\documentclass[conference]{IEEEtran}
\IEEEoverridecommandlockouts
\usepackage{cite}
\usepackage{amsmath,amssymb,amsfonts}
\usepackage{algorithmic}
\usepackage{graphicx}
\usepackage{textcomp}
\usepackage{xcolor}

\usepackage{multirow}
\usepackage{subcaption}
\usepackage{stfloats}

\def\BibTeX{{\rm B\kern-.05em{\sc i\kern-.025em b}\kern-.08em
    T\kern-.1667em\lower.7ex\hbox{E}\kern-.125emX}}
    
\begin{document}

\title{Using Stock Prices as Ground Truth in Sentiment Analysis to Generate Profitable Trading Signals}

\author{\IEEEauthorblockN{Ellie Birbeck}
\IEEEauthorblockA{\textit{Department of Computer Science} \\
\textit{University of Bristol}\\
\textit{Bristol BS8 1UB, UK}\\
eb13817@bristol.ac.uk}
\and
\IEEEauthorblockN{Dave Cliff}
\IEEEauthorblockA{\textit{Department of Computer Science} \\
\textit{University of Bristol}\\
\textit{Bristol BS8 1UB, UK}\\
csdtc@bristol.ac.uk}

}

\maketitle

\begin{abstract}
The increasing availability of ``big" (large volume) social media data has motivated a great deal of research in applying sentiment analysis to predict the movement of prices within financial markets. Previous work in this field investigates how the true sentiment of text (i.e. positive or negative opinions) can be used for financial predictions, based on the assumption that sentiments expressed online are representative of the true market sentiment. Here we consider the converse idea, that using the stock price as the ground-truth in the system may be a better indication of sentiment. Tweets are labelled as Buy or Sell dependent on whether the stock price discussed rose or fell over the following hour, and from this, stock-specific dictionaries are built for individual companies. A Bayesian classifier is used to generate stock predictions, which are input to an automated trading algorithm. Placing 468 trades over a 1 month period yields a return rate of 5.18\%, which annualises to approximately 83\% per annum. This approach performs significantly better than random chance and outperforms two baseline sentiment analysis methods tested. 
\end{abstract}


\section{Introduction}

The field of sentiment analysis is often referred to as ``opinion mining", and from this definition its value is clear: being able to understand not just what a piece of text refers to, but also the attitude towards the text's subject, is a powerful tool. The rise of big data has led to a desire for sentiment analysis to be applied to many areas, and one with obvious potential for significant gain is the financial markets. The ability to accurately read the underlying market sentiment would intuitively suggest an advantage in making and anticipating trading decisions. This premise has motivated much of the research in applying sentiment analysis and machine learning methods in the context of automated trading systems.

One approach to sentiment analysis is text-classification, where predictive models are built by learning from labelled instances of text documents. The need for labelled data is a key barrier in sentiment analysis research, as its context-sensitive nature often requires human evaluation - and even then, humans cannot agree on sentiment around 20\% of the time \cite{human-20-percent}. There are a range of existing sentiment dictionaries which can be obtained from third-party providers, but these usually result in generic scores which are not specific to any domain. In this paper, adpated from \cite{thesis}, we describe a novel approach which labels stock-related text documents according to subsequent changes in the stock price, rather than actual sentiment expressed, and uses this to create and curate dictionaries tailored to individual stocks.

There are many sources of data which can be considered representative of current financial moods. These range from official corporate quarterly reports, through news articles, to chat forums. One such source is Twitter, a globally popular micro-blogging platform which allows its users to publish short messages (tweets) to their followers, and by extension the general public. Since its inception, Twitter has been used by financial investors and speculators to post their trading tips, analysis, and opinions of the markets. This area of activity has increased in recent years, due in large part to the introduction of the \emph{cashtag}. Cashtags are similar to hashtags in that they are metadata labels used to archive tweets of the same tag together, however cashtags exist exclusively for stock tickers. Instead of the \# symbol used to identify hashtags, any stock ticker preceded with a \$ symbol, such as \$AAPL, identifies the tweet as part of the larger conversation about the stock price of the technology company Apple. 
Targeting only tweets containing cashtags allows us to differentiate between casual users tweeting about companies in a consumer capacity, and the community of traders conversing about a stock through the medium of tweets.

Using Twitter to follow the streams of stock-related messages can be thought of as listening to traders shouting across the floor. ``Squawk boxes" were a tool used for this purpose in the past, where intercom speakers allowed the various parties involved in trading decisions to communicate and stay up-to-date on market developments, despite the traders no longer being physically co-located. With trading floors becoming ever more automated, the need for alternative measures of gauging financial moods has become apparent. Twitter provides the large quantities of real-time data required for such a task, but it is important to note that the digital environment is potentially more susceptible to noise, spam, and herd instinct, than the old-fashioned human dynamics of ``open outcry" in the trading pits.

\section{Related Work}

The work in \cite{bollen} produced one of the most widely cited papers using sentiment analysis to predict stock market movements. This investigated correlations between public mood and economic indicators, by measuring collective mood from a small percentage of all tweets in a given time period. Here, the sampling from the entire stream of published tweets takes no regard for topics discussed. Most of the content will therefore be unrelated to what is being predicted, and any stock-specific information in these tweets cannot be specifically inferred as the cause of changing prices. In \cite{info-content-microblogs} this limitation was overcome by using only stock-related tweets, and in particular being the first study to use tweets that contained a specific reference to individual stocks rather than indices or aggregate sentiment. The results of this study showed that the tweets collected did contain valuable information not yet incorporated into market indicators. Similar observations for the need to reduce the scope of information is made by \cite{loughran-mcdonald}, where the common misclassification of sentiment in financial texts was the motivation for constructing a sentiment dictionary specifically tuned to language used in financial literature. 

A major assumption made in all related work that we are familiar with, is that the sentiment expressed in text reflects the true opinions held by the authors, and by extension the true market sentiment. The implications of such an assumption could result in trading decisions being based on information not representative of the true underlying market sentiment. Some works choose to use self-labelled data such as messages posted on StockTwits, a financial communication platform where users can label their posts as `bullish' or `bearish'. However there is evidence of strong biases present in the recommendations made by day traders, in particular with self-disclosed Hold labels actually conveying a positive sentiment rather than neutral \cite{hold-bias-paper}. This claim follows the general optimism of traders, which is further supported by \cite{hope-fear-paper} where the ratio of positive to negative words used in tweets is more than two to one. The optimistic outlook projected by individual traders contrasts that of financial news articles, which often have a negatively skewed bias \cite{media-neg-bias}, and are another source of data for many automated trading algorithms. By using the stock price as the ground-truth, we aim to avoid these biases evident in the labelling of sentiment.

When evaluating model results, \cite{bollen} along with several others performed the testing of their predictions over very small time frames, leading the reliability of results to be questioned. This is noted by \cite{stocktwits-volume-paper}, a study which made 305 predictions over 605 trading days before coming to any conclusions. After this large-scale testing they found no evidence of useful returns from  predictability, although there was evidence of links between trading volume and the number of tweets. The value of contextual features (such as trading volume) in predicting prices is further confirmed by \cite{blog-comm-dynamics}, where a study of the communication dynamics of blogs researched the direction and magnitude of stock price movements in relation to blog comments. Features such as the length, frequency, and response time of comments demonstrated strong correlations with stock market activity. In relation to adding contextual features, there  appears to be promise in adding non-sentiment-based features, such as purely quantitative features. The work in \cite{momentum-trading} modelled a market using two types of trading agent: one which privately observes news but doesn't account for the news observed by other agents; and another which chases trends as the information from news is diffused across the population of traders. This form of `momentum-trading' allows profits to be made from observing only the quantitative measures resulting from under- or over-reaction, not the actual qualitative news content itself.

Although several previous works in this area seem to present reasonable levels of accuracies in predicting stock movements, few test the real value of such predictions: i.e. the ability to generate profit. The work in \cite{elusive-returns} emphasises the challenge of predicting returns, claiming the elusiveness of real returns is due to forecasting models only being sustainable for short periods of time. Many of the works reviewed here have attempted to make relatively long-term predictions, despite the real-time nature of information propagation on Twitter. Unusually, our work in this paper capitalises on the constant stream of news by performing intraday analysis and predicting hourly market movements.

\section{Methodology}

\subsection{Data Collection}
\label{data-collection}
A huge obstacle for many supervised classification tasks is obtaining labelled data - this is particularly true for sentiment analysis, where data often requires manual labelling by humans due to its frequently context-sensitive nature. The price-based approach to labelling sentiment developed here allows us to generate a large data set with little effort, limited only by the number of stock-related tweets available publicly. 

25 stocks were evaluated for their tweet volume and the five with the highest levels of cashtag use were Apple (AAPL), Tesla (TSLA), Twitter (TWTR), Facebook (FB), and Netflix (NFLX). A web-scraping script was used to retrieve a total of 1,474,747 tweets for these  stocks over a 2 year period. Data from 2015 and 2016 were used during training (80\%) and validation (20\%), and data from 2017 held out for testing on a completely new time period. A simple spam-filter targeted the most common form of spam tweet identified, in which a tweet included the cashtags of several different companies, but the content referred to only one or none of those mentioned. Disregarding such tweets by excluding those containing 3 or more cashtags reduced the data set by 23.9\%.

The market data sourced for all stocks contained the date, time, opening price, closing price, high, low, and volume, at one minute intervals from market open to market close. To label the data set of tweets with a classification involved determining the \emph{ground-truth} in terms of the stock price. As previously mentioned, here the classification does not refer to the sentiment expressed in the tweet's content, but is simply an indication of whether or not the stock referred to should be bought or sold, as determined by whether the price rose or fell in the hour following the tweet. 

Temporal information was assigned to each tweet, including the price one hour before and after the tweets, and the volume traded prior to the tweet. Initially, edge cases such as tweets posted in the opening or closing hours of the market, tweets outside of market hours, and on weekends and public holidays, were assigned values through extrapolation. However, this led the data to become noisy as some biases were introduced. For example, one hour of unusual activity before market close would become exaggerated by those values now accounting for multiple hours worth of data. 

\subsection{Language Processing}
To transform the text content of a tweet into a usable object, a tokeniser was applied to parse each tweet, separating them into individual words and filtering to remove irrelevant information. This process included converting characters to lowercase, removing punctuation, reducing three or more concurrently repeated letters to two, removing purely numeric tokens, and replacing URLs with a tag. 



Lemmatisation and stemming processes were not applied, as the reduction of words to their base forms resulted in the loss of valuable information, given the need to analyse each word's predictive power. For example, the words `promised' and `promising' would both generate the lemma `promise', but in reference to a stock performance they could suggest quite different sentiments. The same observation is made in \cite{loughran-mcdonald}, a study involving the creation of a sentiment dictionary attuned to financial contexts, which also considers explicit inflections less prone to errors.

Part-of-speech tagging was also not used in this work, despite its value in many sentiment analysis tasks. Given that our aim was not to identify actual sentiment, but the patterns in language relating to price change, the identification of grammatical categories was deemed less useful. Additionally, the informal language expressed on Twitter produces many words which are not defined as actual words, such as slang terms, abbreviations, and words concatenated for hashtags, which are therefore harder to  tag accurately.

A single matrix of features was created from the corpus of tokens using TF-IDF vectorisation with weight smoothing and L2 normalisation \cite{scikit-learn}. This gives terms occurring frequently in a tweet more weighting, offset if the term also occurs frequently in the whole corpus. This effect of scaling down the weighting for frequent words acts as a filter for generic `stop words' commonly found in a language, such as `and' or `the'. It also results in stop words specific to the corpus being filtered without need for a custom dictionary.

\subsection{Model Development}
\label{model-development}

Three different types of model were evaluated for their  predictive accuracy: Support Vector Machine with RBF kernel; Naive Bayes; and Logistic Regression. All three models were implemented using the scikit-learn library \cite{scikit-learn}. The multinomial variant of Naive Bayes was used, given its suitability with discrete term frequencies, and the model was implemented as standard with Laplace-smoothing and class priors fitted to account for the slight variations in the skewness of training data for each stock. The feature weightings produced were empirical log probabilities indicating how well each word predicts the class of the tweet. The SVM model was initially considered with a variety of kernels (linear, polynomial, Gaussian), but given that the RBF kernel outperformed the rest, it is the only implementation  evaluated fully and comparatively here. For the Logistic Regression model,  a zero-mean Gaussian prior with covariance $\frac{1}{2\lambda} I$ was incorporated for smoothing, along with L2 regularisation used in the penalisation. The model was implemented with a standard minimisation of the following cost function: 

\[ \min_{\theta} \lambda \left \| \theta \right \| ^2 + \sum_{i=1}^n \textup{log}(1+\textup{exp}(-c_i\theta^\top d_i)) \]

For further technical detail see \cite{thesis}. In addition to evaluating the accuracy of each model, two extra performance metrics were considered: the True Buy Rate (TBR); and the True Sell Rate (TSR). These represent the number of correctly predicted Buy/Sell signals divided by the actual number of Buy/Sell signals: essentially a weighted measure of accuracy for each of the classes, without the positive bias which occurs with metrics such as Precision and Recall. These difference between these measures was used to identify occasions where one class dominates the labelled predictions, but this is not evident in the resulting accuracy (for instance, a model always predicting Buy tested on a data set with a majority of Buy labelled tweets will misleadingly suggest good performance).

Our price-based learning approach allows the model to identify words which are uniquely predictive for particular stocks: for example, `timcook' is identified as a Sell word for AAPL. The ability to develop a different model for each stock produces dictionaries which are highly specific, with some words holding opposite sentiments for stocks of rival companies.

\subsection{Time Frame Evaluation}

\begin{figure}[!t]
\centering
\includegraphics[width=0.95\linewidth]{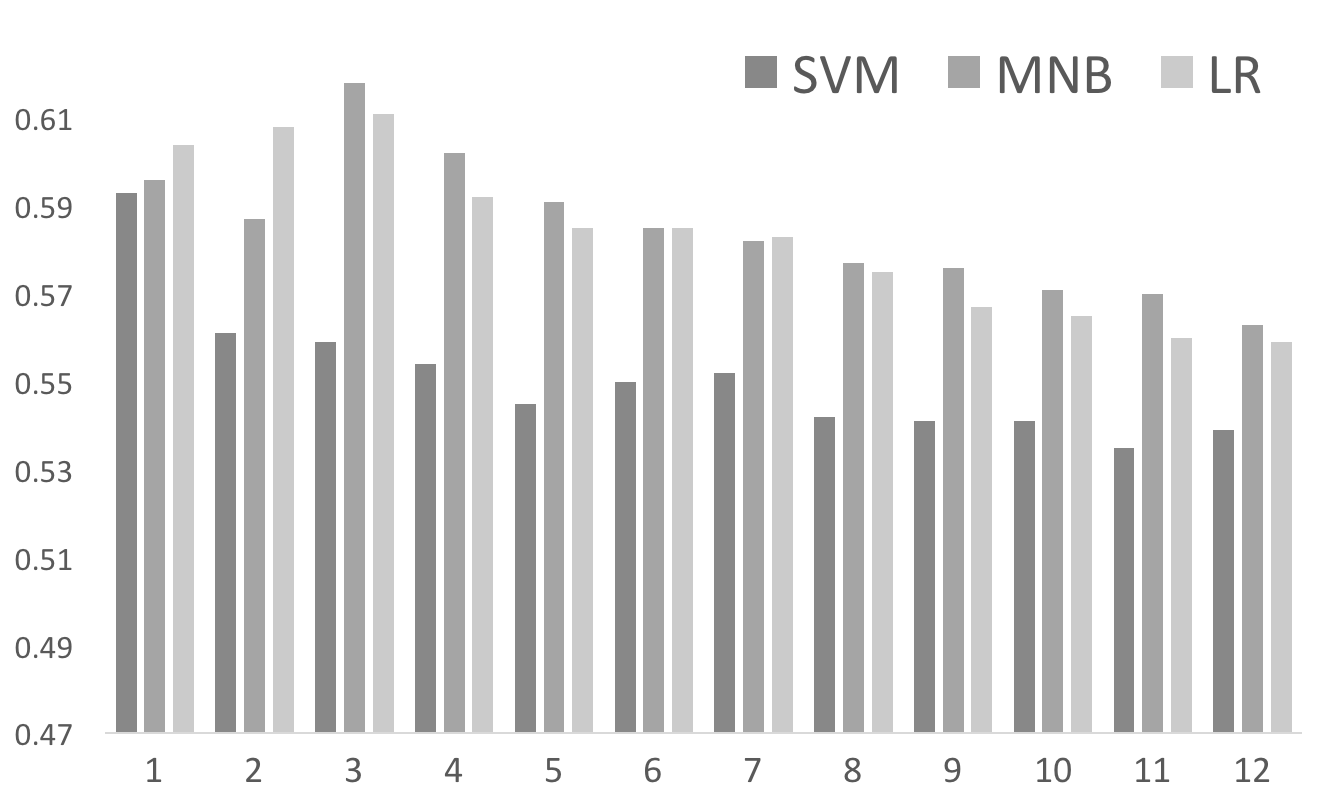}
\caption{Validation set accuracy when trained on increasing time frames, incrementing from 1 month to 12 months.}
\label{fig:month-timeframes}
\end{figure}

One consideration in the model development process was the lifetime of data - for how long is the data collected still relevant? Over two years' of data was initially collected, but it was expected that data from further in the past would be less useful when making predictions for the future. Using classifiers trained on 12 different time-frames in increments of 1 month, revealed a performance peak around 3 months as shown in Fig. \ref{fig:month-timeframes}.

\subsection{Feature Selection}

Text classification often initially results in large feature sets, as the entire collection of words observed in the corpus of documents are considered as features. For example, when training on a sample of 80\% of the AAPL tweets from 2016, the resulting feature vector contained a total of 165,286 features. The filtering steps executed in the tokenisation process reduced this number to under 50,000, but further feature selection was undertaken to choose features based on their statistical significance.

The work in \cite{feature-selection-paper} looked extensively at the effectiveness of feature selection in sentiment classification of tweets. Their results demonstrate the value of using feature selection, and particularly note the choice of ranking system and the size of the feature subset.

\begin{table*}[!t]
\centering
\begin{tabular}{|c|c|c|c|c|c|c|c|c|c|c|c|c|}
\hline
\multirow{2}{*}{Model} & \multirow{2}{*}{Ranker} & \multicolumn{10}{|c|}{Feature Subset Size} \\
\cline{3-12}
 &  & 1000 & 2000 & 3000 & 4000 & 5000 & 6000 & 7000 & 8000 & 9000 & 10000 \\
\hline
\multirow{2}{*}{SVM} & CS & 0.596 & 0.598 & 0.576 & 0.566 & 0.558 & 0.586 & 0.593 & 0.558 & 0.562 & 0.564 \\
 & FV & 0.595 & 0.569 & 0.597 & 0.596 & 0.604 & 0.598 & 0.597 & 0.590 & 0.584 & 0.558 \\
\hline
\multirow{4}{*}{MNB} & CS & 0.628 & 0.649 & 0.649 & 0.650 & \textbf{0.651} & 0.648 & 0.642 & 0.641 & 0.642 & 0.629 \\
 & FV & 0.627 & 0.648 & 0.646 & 0.649 & 0.639 & 0.628 & 0.641 & 0.641 & 0.640 & 0.635 \\
 & MI & 0.511 & 0.517 & 0.529 & 0.530 & 0.531 & 0.542 & 0.531 & 0.526 & 0.523 & 0.527 \\
 & RFE & 0.510 & 0.511 & 0.508 & 0.511 & 0.514 & 0.517 & 0.516 & 0.518 & 0.522 & 0.525 \\
\hline
\multirow{4}{*}{LR} & CS & 0.629 & 0.629 & 0.630 & 0.629 & 0.625 & 0.624 & 0.616 & 0.613 & 0.609 & 0.607 \\
 & FV & 0.616 & 0.621 & 0.615 & 0.616 & 0.616 & 0.613 & 0.614 & 0.613 & 0.612 & 0.601 \\
 & MI & 0.515 & 0.521 & 0.525 & 0.520 & 0.522 & 0.528 & 0.524 & 0.525 & 0.525 & 0.524 \\
 & RFE & 0.521 & 0.526 & 0.526 & 0.530 & 0.527 & 0.531 & 0.527 & 0.526 & 0.528 & 0.526 \\
\hline
\end{tabular}
\caption{Validation set accuracy of each model with varying feature selection methods and subset sizes. In the `Model' column, `SVM' refers to Support Vector Machine, `MNB' refers to Multinomial Naive Bayes, `LR' refers to Logistic Regression. In the `Ranker' column, `CS' refers to Chi-Squared, `FV' refers to F-value, `MI' refers to Mutual Information, `RFE' refers to Recursive Feature Elimination. The best accuracy is 0.651 (in bold) for MNB with CS at size 5000. See text for further discussion.}
\label{table:feature-subsets}
\end{table*}

\subsubsection{Chi-Squared}

The ${\chi}^2$ test is a very well-known non-parametric test to determine whether two events are independent, and this can be applied to feature selection by thinking of the two events as term occurrence and class occurrence. In this context, the ${\chi}^2$ feature selection is calculating whether a word occurring in a tweet is independent of whether that tweet is classified as Buy or Sell. Words are ranked according to their value as calculated by:

\[ {\chi}^2(d,t,c) = \sum_{e_t \in \{0,1\}} \sum_{e_c \in \{0,1\}} \frac{(N_{e_t e_c} - E_{e_t e_c} )^2}{E_{e_t e_c}} \]

\noindent
where $d$, $t$, and $c$ refer to document, term, and class respectively, $N$ is the observed frequency in $d$, $E$ is the expected frequency in $d$, $e_c=1$ if the document is in class $c$ and $0$ if not, and $e_t=1$ if the document contains term $t$ and $0$ if not. For example $N_{e_t=1 e_c=1}$ represents the observed frequency of term $t$ occurring in document $d$ which is of class $c$. 

If the events are dependent (and therefore the classification of Buy or Sell depends on the occurrence of the word), then this signifies that the word is useful and should be included as a feature. All words in the tweet corpus for each stock are ranked according to the ${\chi}^2$ statistic, and only the highest ranking words are kept in the feature vector for that stock. 

\subsubsection{ANOVA F-value}

Analysis of Variance (ANOVA) refers to a group of parametric statistical models and tests which calculate the variation between and within groups, and one of the key elements computed in ANOVA statistics is the F-value, the ratio:

\[ F = \frac{\textup{variance between groups}}{\textup{variance within groups}} \]

\noindent
The F-value is used to estimate the linear dependency between two variables (here this refers to the class and the term of a document), and as with ${\chi}^2$, this method of feature selection returns univariate scores for the features which can be used for ranking the features in order of their value in terms of classifying new instances.

\subsubsection{Mutual Information}

Whereas the ANOVA $F$-value test estimates the degree of linear dependence between events, Mutual Information is a measure of statistical dependency in any form. Generally, it measures the amount of information known about one event through knowledge of another event. In this context, it quantifies the amount of information regarding the class of a tweet that is gained through observation of the word within that tweet. The statistic is calculated as:

\[ MI(d, t,c)=\sum _{e_t \in \{0,1\}}\sum _{e_c \in \{0,1\}}p(e_t,e_c)\log {\left({\frac {p(e_t,e_c)}{p(e_t)\,p(e_c)}}\right)} \]

\noindent
where $d$, $t$, and $c$ again refer to document, term, and class respectively, $p(e_t,e_c)$ refers to the joint probability distribution of $e_t$ and $e_c$, and $p(e_t)$ and $p(e_c)$ refer to their individual marginal probability distributions.

\subsubsection{Recursive Feature Elimination}

A non-statistical approach to feature selection was also considered, whereby an optimal subset of features is defined by recursively selecting fewer and fewer features, gradually pruning those that have the lowest contribution in the current subset. The motivation behind weight-based feature selection methods such as this, is to justify the value of a feature based on the error rate resulting from its removal from the set \cite{feature-extraction-book}. 

This method was not applicable to the SVM classifier developed here, because the mapping function of the RBF kernel is not explicitly known and therefore the weight vector required for recursive feature selection cannot be determined \cite{rfe-rbf-paper}. Additionally, the Mutual Information feature selection method in combination with the SVM model incurred extremely long run-times and lower performance in initial testing, so was not evaluated fully and is therefore also not included in the results, which are displayed in Table \ref{table:feature-subsets}.

The highest accuracy was achieved using Multinomial Naive Bayes with the top 5000 features rated by ${\chi}^2$ rankings. The use of ranking methods over other methods of dimensionality reduction gives the key advantage of being able to identify which words are contributing most towards the classification.

\subsection{Stock-Related Feature Construction}

In addition to the feature vector of dictionary words, three aspects relating to the quantitative stock performance were considered: the previous direction of the stock price; the volume of stock transactions; and the temporal relationship of stock price. 

As directional price trends are expected to continue, a feature representing a previous bullish or bearish trend was added. This was structured as a dense matrix of binary values and transposed then converted to sparse for concatenation with the existing matrix of word features. For further detail see \cite{thesis}.

A feature for trading volume represented the total trading volume of the stock during the hour prior to the tweet being posted. This was tested as both an integer value and a binary value (based on whether or not the integer value exceeded a threshold of the average volume traded per hour - used to achieve linear separability).

\begin{figure}[!t]
\centering
\begin{subfigure}{.5\linewidth}
  \includegraphics[width=\linewidth]{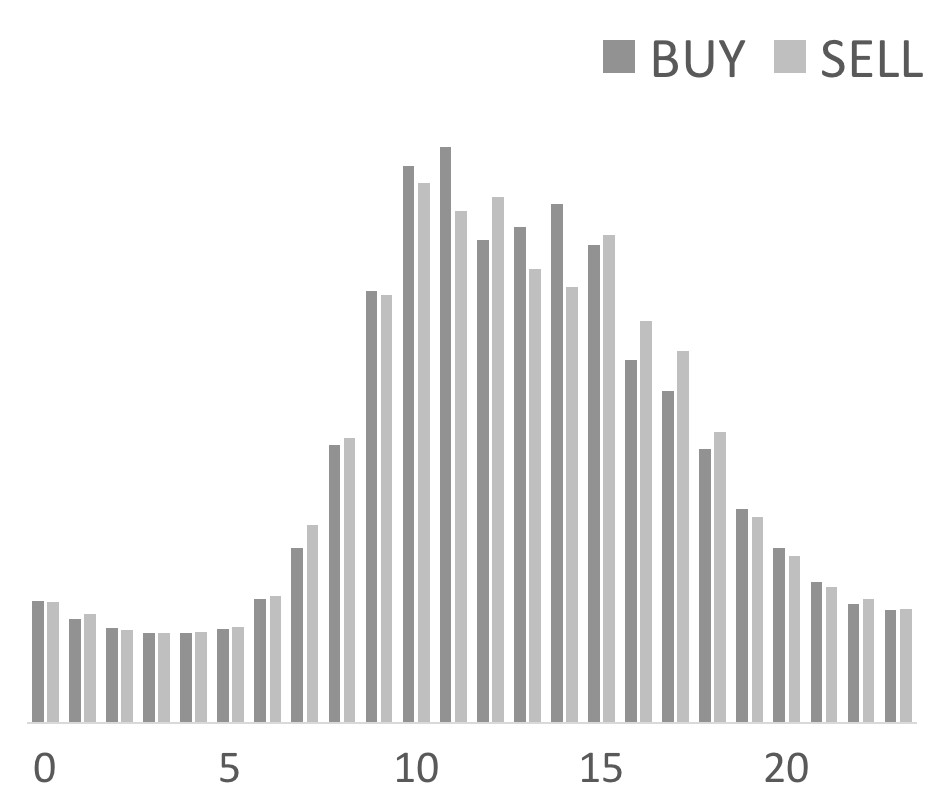}
  \caption{Hourly Trends}
  \label{fig:sub1}
\end{subfigure}%
\begin{subfigure}{.5\linewidth}
  \includegraphics[width=\linewidth]{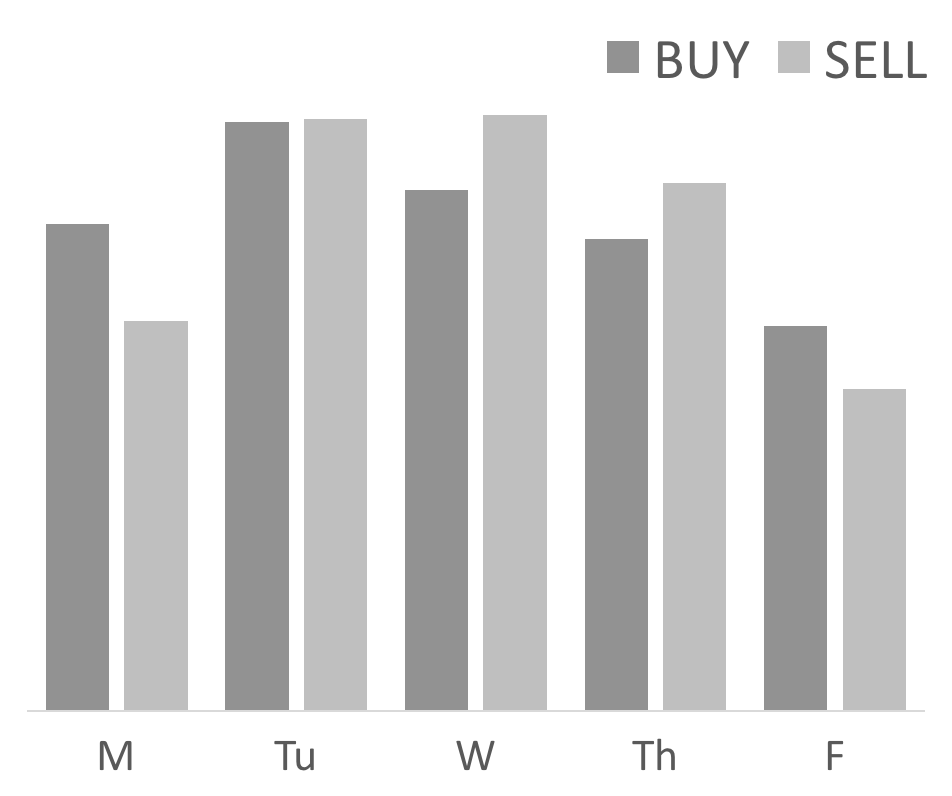}
  \caption{Daily Trends}
  \label{fig:sub2}
\end{subfigure}
\caption{Histograms showing temporal trends of the frequencies of directional price signals.}
\label{fig:hist-temporal-trends}
\end{figure}

To test whether stock price fluctuations were correlated with time, trends across hourly, daily, monthly, and quarterly time frames were evaluated. There was an insufficient quantity of monthly and quarterly periods to extrapolate the patterns observed, but analysis of the hourly and daily frequency distributions shown in Fig. \ref{fig:hist-temporal-trends} demonstrated that a feature for weekday could represent the differing distributions.

\section{Results}

Theoretically any predictive accuracy result above 50\% is promising in the context of trading, as it defies the efficient market hypothesis by performing better than random chance. However, to evaluate the real value of the model, its predictions need to be tested in terms of their ability to generate a profit.

\subsection{Profitability of trading algorithm}
\label{profitability-section}

A simple trading algorithm was implemented whereby the total data set of tweets for January 2017 acts as input and thereafter, as if running live, a decision is processed every hour from 10am to 3pm each trading day. All tweets posted throughout the previous hour are analysed for their sentiment, and a 50\% threshold determines whether to buy or sell the stock. The position is then held for an hour, before either a profit or loss is taken. Given the 6 trades placed per trading day, and during the testing period of January 2017 markets being open for trading on 20 days, a total of 120 trades were executed per stock.

The trade placed each hour has a size of 100 shares, for the sake of simplicity and consistency with regard to trade execution - despite different stocks having different values and the values of each stock changing over time. To provide transparency with regard to the profitability of the algorithm, the results are presented initially without any fees incurred, so that the total amount gained or lost is entirely the result of the predictions. 

To compare the resulting profits for each stock, a percentage of returns rather than absolute value is calculated. This is achieved by first determining an initial account size required to trade at such volumes (100 shares) for the specific stock, allowing for a negative margin of 10\% of the account value in order to hold potential losses. For example, given the maximum price per share of \$122 for Apple stock in January 2017, the required account size is calculated to be \$13,420. The trades placed for Apple had a 64.1\% success rate, with 77 out of 120 generating a profit, giving a total gain of \$729.50. This monthly return rate is therefore 5.44\%. The value could be extrapolated to a more widely recognised annualised return rate of 88.6\%, although the cumulative effect of compounding one result in this way gives a more uncertain value, as the likelihood of maintaining this exact monthly rate throughout the year is low. 

Table \ref{table:monthly-return-rate} displays the monthly return rates for trading Apple, Tesla, Twitter, and Facebook stock. Data was also collected for Netflix, but at the time of testing we decided that the quantity of tweets referencing the stock was not sufficient to use in the trading algorithm, as automated trading decisions were being made on the basis of single tweets in a significant number of cases. 

The study in \cite{trading-costs} used fund portfolio holdings and transaction data to investigate trading costs. Using the reported per unit costs of commissions, bid-ask spread, and price impact, for large-cap growth fund groups, an estimation of the total per unit trading cost is 0.48\%. Given our portfolio turnover rate of 100\% (all shares are traded), and the replacement of transactions (Buy orders are followed by Sell orders after 1 hour and vice versa), the monthly trading cost amounts to 0.96\% of the account size. Annual expenditure trading costs across all fund groups in \cite{trading-costs} is estimated at 1.44\%. Taking these estimated fees into account diminishes the profits of trading Twitter stock to almost zero, and for the remaining 3 stocks the return rates are reduced but similar in outcome. The results including the monthly costs for each stock are shown in Table \ref{table:monthly-return-rate-with-fees}.

\begin{table}[!t]
\centering
\begin{tabular}{|c|c|c|c|c|}
\hline
{\textbf{Stock}} & {AAPL} & {TSLA} & {TWTR} & {FB} \\

\hline
{\textbf{Return}} & $5.44\% $ & $9.68\%  $ & $1.0\% $ & $-3.12\%  $ \\

\hline
\end{tabular}
\caption{Monthly return rate from January 2017.}
\label{table:monthly-return-rate}
\end{table}

\begin{table}[!t]
\centering
\begin{tabular}{|c|c|c|c|c|}
\hline
{\textbf{Stock}} & {AAPL} & {TSLA} & {TWTR} & {FB} \\

\hline
{\textbf{Return}} & $4.48\% $ & $8.72\%  $ & $0.04\% $ & $-4.08\%  $ \\

\hline
\end{tabular}
\caption{Monthly return rate from January 2017 with estimated fees incurred.}
\label{table:monthly-return-rate-with-fees}
\end{table}

Table \ref{table:trade-breakdown} displays a breakdown of the trades placed, and it is interesting to note the majority of orders were Sell, despite all four stocks rising in price over the month tested. The case of Twitter's hugely negative skew is likely the result of two large price drops in the training period generating lots of negative data. Also evident is a relationship between the quantity of tweets and the profitability of the algorithm, as the stocks are listed in order of liquidity of cashtag use during the testing period (AAPL, TSLA, TWTR, FB), and the profits decrease correspondingly. The number of trades placed for Facebook actually reduced by 10\% purely due to an insufficient number of tweets (zero) to generate a signal in some hours. For further discussion see \cite{thesis}.

\begin{table}[!t]
\centering
\begin{tabular}{|c|c|c|c|c|c|}
\hline
Stock & Order & \multicolumn{2}{|c|}{Orders Placed} & \multicolumn{2}{|c|}{Orders Correct} \\
\hline
\multirow{2}{*}{AAPL} & Buy & 51 & 42.5\% & 32 & 62.7\% \\
& Sell & 69 & 57.5\% & 45 & 65.2\% \\
\hline
\multirow{2}{*}{TSLA} & Buy & 79 & 65.8\% & 45 & 60.0\% \\
& Sell & 41 & 34.2\% & 23 & 56.1\% \\
\hline
\multirow{2}{*}{TWTR} & Buy & 9 & 7.5\% & 6 & 66.7\% \\
& Sell & 111 & 92.5\% & 52 & 46.8\% \\
\hline
\multirow{2}{*}{FB} & Buy & 23 & 21.3\% & 15 & 65.2\% \\
& Sell & 85 & 78.7\% & 35 & 41.1\% \\
\hline
\end{tabular}
\caption{Breakdown of orders executed for each stock.}
\label{table:trade-breakdown}
\end{table}

The web-scraping method used for data collection is a significant barrier to further testing as it does not return all tweets posted, and the limitations imposed by the Twitter API prevent free access to the full data set of posted tweets. However given that the accuracy of tweet predictions is $\geq50\%$ across all stocks, the results indicate that larger quantities of tweets per hour could certainly be a contributing factor towards higher profitability in the resulting trading algorithm.

\subsection{Significance of results}

The study by \cite{bollen} which made daily directional predictions based on the correlation between the emotion `calm' and the movements of the Dow Jones Industrial Average (resulting in the ``Twitter Hedge Fund"), tested the statistical significance of their result occurring by chance using a model based on the binomial distribution.

The same assessment method is applied to the results in this paper. Using the count of 253 correct trades placed out of a total 468, with the same 50\% chance of success on each trade, gives a probability of 0.789\% for achieving this result by chance. As testing was performed on 20 trading days out of the total 85 day period, the approximate number of time frames for selection equals 4.25, and the likelihood of the probability holding for a random period of such time by chance is calculated to be 3.35\% - a similar result to \cite{bollen} which as they state, means the accuracy is most likely not due to chance or favourable test period selection.

The cumulative binomial probability can also be calculated, which instead of giving the likelihood of the exact outcome resulting from the 468 trades, which seems an overly precise constraint, gives the probability that at least 253 of the 468 trades were correct. In other words, what is the chance of a trading algorithm performing equal to \emph{or better than} the one produced here by chance. This result is 3.57\%. When applying this value in combination with the likelihood of selecting a favourable testing period, the probability rises to 15.2\%.
Although this probability is still low, it is not negligible, and so given the fact that all trades were placed in January 2017, one additional month of data for AAPL was tested in December 2016 (using the previous 3 months to train the classifier), for an added level of validation that the chosen time period was not a compromising factor in the credibility of the results produced. The resulting profitability of the algorithm for this month gives a return rate of 3.86\%. This is not as high as the previous test period, but the evidence of further profitability acts to mitigate concerns regarding selective time periods. 

\subsection{Comparison against baseline methods}

Two methods of sentiment-analysis-based trading approaches were evaluated as baseline measures. The stock-specific price-based approach developed in this paper is from here on referred to as Method A. Method B uses the popular existing sentiment dictionary SentiWords \cite{sentiwords} based on research in \cite{senti-research}, to give tweets a generic rating of positive, negative, or neutral. Method C does the same, but using the Loughran \& McDonald dictionary \cite{loughran-mcdonald}, developed for use in the financial domain.

Applying Method B generated a highly positive skew in the classification of tweets (85\%), further supporting the claims in \cite{loughran-mcdonald} that applying generic sentiment dictionaries gives inaccurate results due to the high number of words which are generally viewed as positive, but in a financial context are deemed neutral. Method C had a negatively skewed dictionary, but an almost even classification of tweets. Despite this, the low total number of words in the dictionary results in many tweets processed as unclassified, and not contributing towards a trading signal.

\begin{figure}[!t]
\centering
\includegraphics[width=0.9\linewidth]{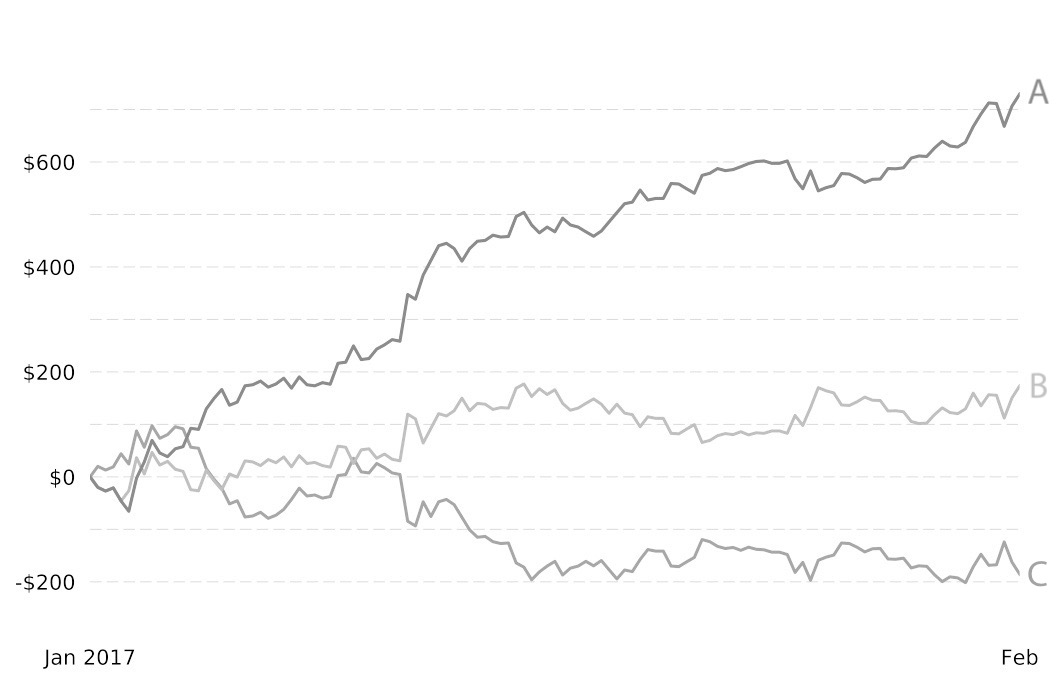}
\caption{Profits trading AAPL stock with Methods A, B, C.}
\label{fig:aapl-comparison}
\end{figure}

\begin{figure}[!t]
\centering
\includegraphics[width=0.9\linewidth]{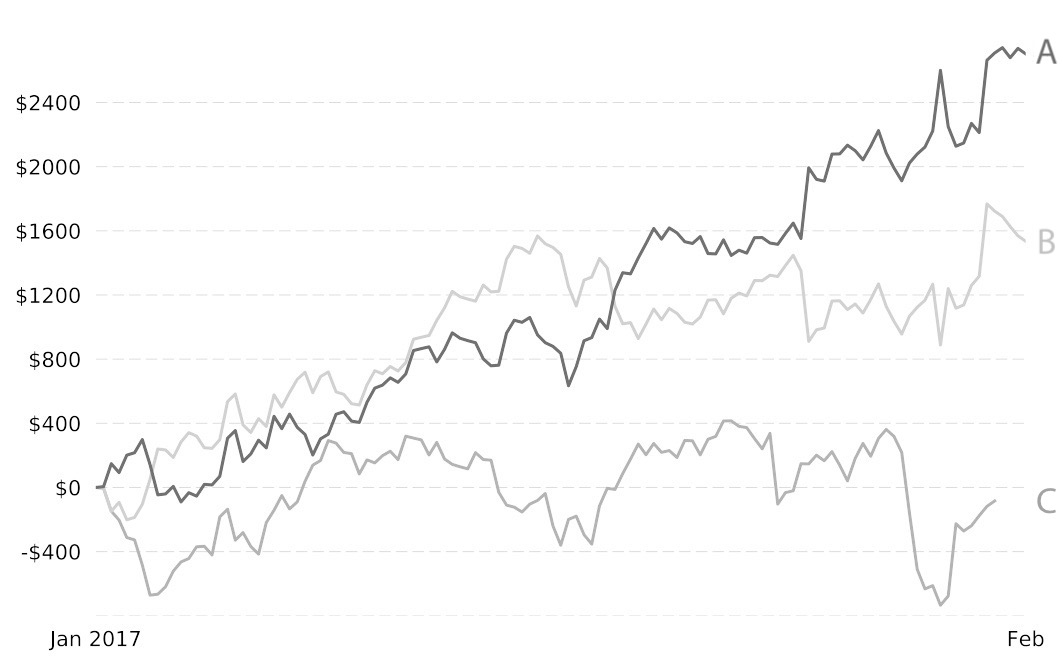}
\caption{Profits trading TSLA stock with Methods A, B, C.}
\label{fig:tsla-comparison}
\end{figure}

\begin{figure}[!t]
\centering
\includegraphics[width=0.9\linewidth]{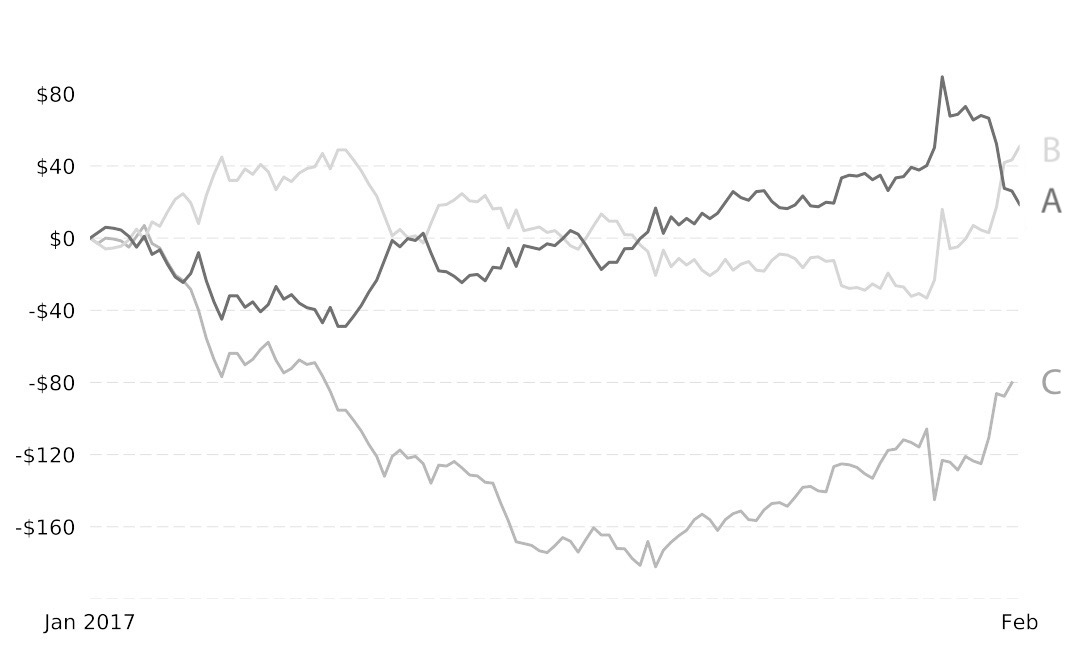}
\caption{Profits trading TWTR stock with Methods A, B, C.}
\label{fig:twtr-comparison}
\end{figure}

\begin{figure}[!t]
\centering
\includegraphics[width=0.9\linewidth]{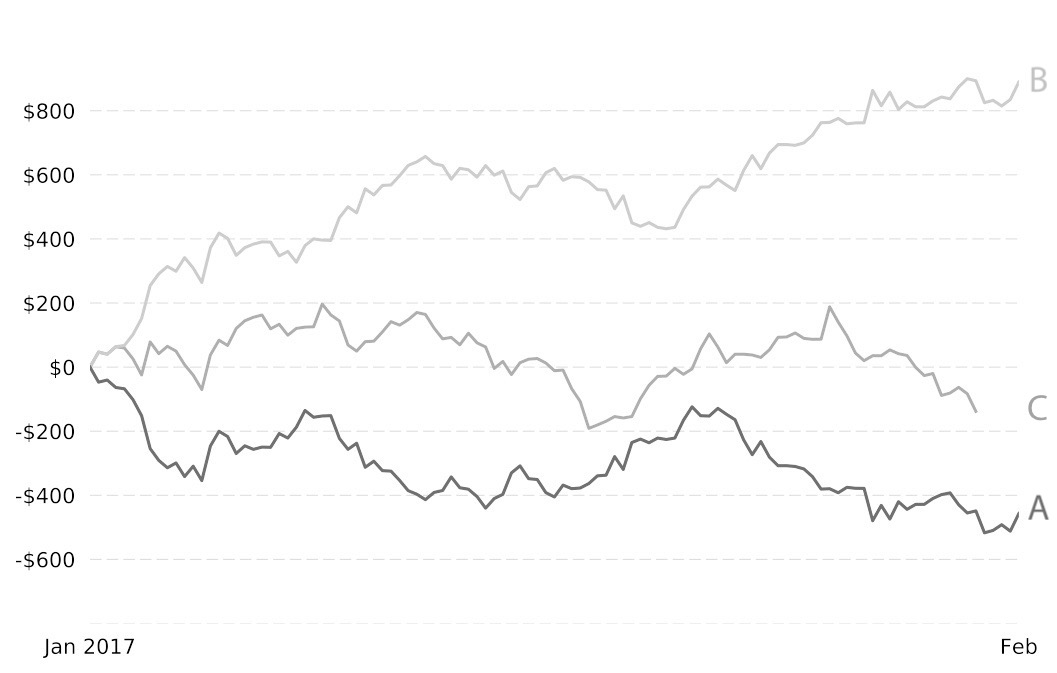}
\caption{Profits trading FB stock with Methods A, B, C.}
\label{fig:fb-comparison}
\end{figure}

Figs \ref{fig:aapl-comparison}-\ref{fig:fb-comparison} display a comparison of the profits of each method. Although Method B appears consistently profitable, it cannot be considered a good method in practice. The high positive skew essentially produces a `buy-and-hold' strategy, with 99\% of the automated trades placed as Buy orders, and it is coincidental that this performs well for the testing period. This problem can be identified by analysing the difference between the True Buy Rate and True Sell Rate as discussed in Section \ref{model-development}, which is ideally near the minimum of 0, but in this instance is near the maximum of 100.

\subsection{Sharpe Ratio of returns}

The Sharpe ratio is a method of measuring the performance of an investment in relation to its risk. This measure computes the expected return of an investment, or in this case the use of a trading strategy, per unit of risk \cite{sharpe-ratio}. The calculation is as follows:

\[ S = \frac{\bar{d}}{\sigma_d } = \frac{E[R_i - R_b]}{\sqrt{\textup{var}[R_i - R_b]}}\]

where $\bar{d}$ represents the differential return and $\sigma_d$ represents the standard deviation of $d$, and $E[R_i - R_b]$ represents the expected return on investment $i$, $R_i$, compared to the return on a benchmark $b$, $R_b$. This particular expression is the version redefined by the author as the {\em ex-ante} Sharpe ratio. 

Here the benchmark chosen for comparison is the S\&P500, a U.S. stock market index composed of 500 large companies, which is widely considered a good representation of the U.S. stock market and economy in general. Each individual stock return rate is compared to the benchmark return of investment in the S\&P500 using a `buy-and-hold' strategy, which involves buying the index at the beginning of the month and holding this position throughout, generating a profit equal to the total increase or decrease in price. 

The measure allows investments or trading strategies to be compared on a risk-adjusted return basis, meaning that those with similar return rates can be ranked in terms of which offer a higher return per unit of risk. The resulting Sharpe ratios displayed in Table \ref{table:sharpe-ratio} are further indicators of return performance for each stock. 

\begin{table}[!t]
\centering
\begin{tabular}{|c|c|}
\hline
Stock & Sharpe Ratio \\
\hline
AAPL & 2.78  \\
TSLA & 3.06 \\
TWTR & -0.016 \\
FB & -5.46 \\
\hline
\end{tabular}
\caption{Sharpe ratios for the return profit of each stock against a benchmark `buy-and-hold' strategy for the S\&P500.}
\label{table:sharpe-ratio}
\end{table}

\section{Discussion}

Our work investigated whether using the stock price to label stock-related tweets could provide a better indication of financial sentiment than methods used in current practice, and tested the ability of such an approach to generate real profits in an automated trading system. 

The results show that there is value in this idea - the creation of stock-specific dictionaries for individual companies along with basic quantitative measures relating to stock performance produce a classifier that can label tweets with accuracies consistently above the 50\% baseline for random guessing. When using these predictions in a trading system, the execution of 468 trades over a 1 month period generates a total return rate of 5.18\%. The real-time nature of information propagation on Twitter is used to our advantage, with the hourly execution of trades allowing the system to capitalise on frequent changes in price, and yields much higher potential profits than when predicting longer trends. Compared to two existing sentiment analysis methods tested, our approach described here outperforms these baseline measures.

Given the limitation of the evaluation to only 4 stocks, there would be great value in further testing with access to the full public domain of tweets. The likely correlation between quantity of tweets per hour and the profitability of trading decisions indicate that, as mentioned in Section \ref{profitability-section}, obtaining full access to published tweets could increase return rates for the stocks evaluated. However the low volume of stock-related tweets in general is the bigger problem, with the cashtags for many less newsworthy companies not mentioned on a sub-hourly basis.

\section{Further Work}

There are two main areas that we intend to explore in future work. Firstly we aim to introduce aspect-level sentiment analysis in order to improve the quality of data used, through filtering for future-oriented tweets which exclusively refer to the intended stock. Aspect-level sentiment analysis could also be used to analyse joint company mentions in order to predict stock co-movement.

We also aim to develop a more advanced trading system. This could be achieved at a minimum by adding a third class with a neutral decision of Hold. Additionally, different percentage thresholds could be used for placing orders, to increase the accuracy of decisions but potentially reducing the total number of orders placed. We also aim to investigate evolving the classifier beyond simply directional predictions to a points-based system.

\section*{Acknowledgment}

This work is based on E. Birbeck's Master's thesis \cite{thesis}, which won the 2017 University of Bristol Bloomberg Prize for Best Final Year Project in Machine Learning; we are grateful to Bloomberg for their generosity and recognition.

\bibliographystyle{ieeeconf}

\end{document}